\documentclass[preprintnumbers, floatfix, preprintnumbers, letterpaper, superscriptaddress,nofootinbib]{revtex4}
\usepackage{graphicx}
\usepackage{microtype}
\usepackage{amsmath}
\usepackage{amssymb}
\usepackage{subfigure}
\usepackage{hyperref}
\usepackage{url}
\usepackage{xcolor}
\usepackage{color}
\usepackage{mathrsfs}
\usepackage{calrsfs}
\usepackage{amsfonts}
\usepackage{eufrak}
\usepackage{tabularx}
\usepackage{eucal}
\usepackage{latexsym}
\usepackage{ragged2e}
\usepackage{epsfig}
\usepackage{textcomp}

\usepackage{caption}
\DeclareCaptionJustification{justified}{\leftskip=0pt \rightskip=0pt \parfillskip=0pt plus 1fil}
\definecolor{vividviolet}{rgb}{0.62, 0.0, 1.0}
\definecolor{amaranth}{rgb}{0.9, 0.17, 0.31}
\definecolor{palatinateblue}{rgb}{0.15, 0.23, 0.89}
\definecolor{brightpink}{rgb}{1.0, 0.0, 0.5}
\definecolor{cornflowerblue}{rgb}{0.39, 0.58, 0.93}
\definecolor{deepcarminepink}{rgb}{0.94, 0.19, 0.22}
\definecolor{radicalred}{rgb}{1.0, 0.21, 0.37}

\hypersetup{ linktoc=all,
    colorlinks, linkcolor={palatinateblue},
    citecolor={brightpink}, urlcolor={amaranth}
}

\graphicspath{{Images/}}

\renewcommand{\d}[1]{\ensuremath{\operatorname{d}\!{#1}}}

\graphicspath{{Images/}}
\renewcommand{\d}[1]{\ensuremath{\operatorname{d}\!{#1}}}
\makeatletter
\def\@fnsymbol#1{\ensuremath{\ifcase#1\or \ddagger \or  $\textleaf$  \or \dagger
\else\@ctrerr\fi}}%

\makeatother

\def\sideremark#1{\ifvmode\leavevmode\fi\vadjust{\vbox to0pt{\vss
 \hbox to 0pt{\hskip\hsize\hskip1em
 \vbox{\hsize1.3cm\tiny\raggedright\pretolerance10000
 \noindent #1\hfill}\hss}\vbox to8pt{\vfil}\vss}}}%
\def\beq{\begin{equation}}
\def\eeq{\end{equation}}

\begin{document}

\title{Curvature Invariants and Lower Dimensional Black Hole Horizons}

\author{Daniele \surname{Gregoris}}
\email{danielegregoris@libero.it}
\affiliation{Center for Gravitation and Cosmology, College of Physical Science and Technology, Yangzhou University, \\180 Siwangting Road, Yangzhou City, Jiangsu Province, P.R. China 225002}
\affiliation{School of Aeronautics and Astronautics, Shanghai Jiao Tong University, Shanghai 200240, China}

\author{Yen Chin \surname{Ong}}
\email{ycong@yzu.edu.cn}
\affiliation{Center for Gravitation and Cosmology, College of Physical Science and Technology, Yangzhou University, \\180 Siwangting Road, Yangzhou City, Jiangsu Province, P.R. China 225002}
\affiliation{School of Aeronautics and Astronautics, Shanghai Jiao Tong University, Shanghai 200240, China}

\author{Bin \surname{Wang}}
\email{wang_b@sjtu.edu.cn}
\affiliation{Center for Gravitation and Cosmology, College of Physical Science and Technology, Yangzhou University, \\180 Siwangting Road, Yangzhou City, Jiangsu Province, P.R. China 225002}
\affiliation{School of Aeronautics and Astronautics, Shanghai Jiao Tong University, Shanghai 200240, China}

\begin{abstract}
It is known that the event horizon of a black hole can often be identified from the zeroes of some curvature invariants. The situation in lower dimensions has not been thoroughly clarified. 
In this work we investigate both (2+1)- and (1+1)-dimensional black hole horizons of static, stationary and dynamical black holes, identified with the zeroes of scalar polynomial and Cartan curvature invariants, with the purpose of discriminating the different roles played by the Weyl and Riemann curvature tensors. The situations and applicability of the methods are found to be quite different from that in 4-dimensional spacetime.   The suitable Cartan invariants employed for detecting the horizon can be interpreted as a local extremum of the tidal force suggesting that the horizon of a black hole  is a genuine special hypersurface within the full manifold, contrary to the usual claim that there is nothing special at the horizon, which is said to be a consequence of the equivalence principle.
\end{abstract}

\maketitle

\section{Introduction: Locating the Event Horizon of a Black Hole}

 Black holes have been defined differently across different disciplines like astrophysics, analogue gravity and mathematical relativity \cite{defbh}. For example section 12.4 of \cite{wald} introduces black holes as regions of spacetime from which nothing -- not even light -- can escape from, making them  some of the most studied solutions to the Einstein equations of general relativity \cite{exact}. After a few decades from their introduction as a pure mathematical concept,  an interpretation as astrophysical object which  -- as far as we know -- can be fully characterized in terms of their mass, electric charge and angular  momentum  has been proposed.   Since techniques for estimating the values of these parameters became available (for example through the keplerian motion of orbiting stars \cite{keplerian}), they  must be regarded as physical quantities and not only as mathematical oddities entering the theory equations.  The existence of black holes   in the actual   Universe we live in has been indirectly established through the study of the X-rays emitted by the matter falling into them before crossing the horizon \cite{nature1,ruffini}. Much more recently,  the direct detection of gravitational waves by the LIGO collaboration has brought to the first observational identification of black hole binary systems in the Universe \cite{ligo1}. However the prediction of the energy spectra of the waves emitted by coalescing black holes is almost entirely based on numerical techniques  \cite{num1,num2,num3}  and analytical methods are still scarce according to the latest review on the experimental and theoretical methods for studying gravitational waves \cite{natgw}.   From the theoretical point of view it has been argued that the evolution of a black hole can be fully accounted for by four mechanical  (or also known as  ``thermodynamical" \cite{texas})  laws which relate its mass, electric charge, angular  momentum, area of the horizon, surface gravity and temperature \cite{poisson}. Note that the notion of the horizon is crucial in these laws: if, eventually, we have the technology to test these laws in the actual Universe, we would need to first correctly identify the horizon of a black hole.

Moreover, cornerstone theorems by Hawking and Penrose proved that physical singularities (not to be confused with coordinate singularities which can be removed by a change of the coordinate system) can occur in the theory of classical general relativity \cite{haw1,haw2,haw3,haw4}. These results have been checked explicitly with respect to the spacelike singularity  that occurs  in non-rotating black hole with no electric charge,   and timelike singularities for rotating and electrically charged black holes,  exploiting the knowledge of such exact analytical solutions. Spacetimes admitting a physical singularity are geodesically incomplete, meaning that timelike and/or lightlike geodesics cannot be extended after reaching the singular point. Thus it is not possible to know the physical behavior of such singularities, and in the case  in which  they are naked (that is they can be seen by   an observer located at the future null infinity), physical theories would lose predictability. Thus a mechanism  that prevents   observers to directly see singularities is needed in a physically solid formulation of  general relativity. The  presence of an event horizon   which hides curvature singularities seems the most convenient proposal to address this puzzle: this is the so-called cosmic censorship conjecture \cite{censor}. Unfortunately it has been claimed that  even   in principle   at least   in the near future it will not be possible to detect  black hole horizons by direct astrophysical measurements   because  indirect measurements   can only claim the existence of  a  black hole by measuring its mass, which must be higher than the mass of a neutron star emitting the same X-rays waves, and its accretion disk \cite{imp}.   Moreover,  the theoretical interpretation of the shadow imaging of a black hole observed by the Event Horizon Telescope has been done only with computer simulations \cite{eht}.    Therefore at this stage, we must first focus on the  analytical  study of black hole horizon, and the localization of it becomes an issue of crucial importance. 

On the other hand, modifications to the Hilbert-Einstein action  have been proposed in  the   recent literature for many reasons ranging from the modeling of the late time accelerated expansion of the Universe \cite{capo}, to the analysis of its first inflationary stages \cite{staro}, to the modeling of the rotation curve of galaxies \cite{tor}, just to mention a few examples.  Lower dimensional -- both (2+1)- and (1+1)-dimensional -- theories for gravity have been formulated with a twofold purpose of clarifying the mathematical properties of the Einstein equations (since they are a system of coupled nonlinear partial derivative equations,  their solution has been derived so far only in simplified cases requiring strong symmetry assumptions)  in an analogue but formally simpler case, and for developing  possible frameworks for a quantum formulation of the gravitational interaction \cite{witten1, witten2, 1plus1, sc1plus1}. Even though lower-dimensional gravity is a pure topological theory, it admits black hole solutions, the first example was found in (2+1) dimensions in the presence of a negative cosmological constant \cite{zanelli1, zanelli2}. It is therefore of interest to see whether the mechanical and thermodynamical laws governing the evolution of (3+1)-dimensional black holes hold also for the lower-dimensional case. Assuming  that the  thoughts leading to the cosmic censorship conjecture still apply, it is of crucial importance for this goal to correctly identify the black hole horizon.

 Besides, black holes are an important tool in analogue gravity  for deepening the understanding of physical systems which seem completely unrelated to them at first sight \cite{analogue}. For example their acoustic or sonic properties are the exact counterpart to their optical ones in astrophysics, that is they admit a surface interpretable as an event horizon since sound signals cannot escape from it, which can be determined by the fact that the speed of flow of the fluid inside which they are formed becomes smaller than the speed of sound within the fluid. More recently an experimental detection of a black hole horizon has been proposed as realizable in a laboratory setting exploiting a Bose-Einstein condensate \cite{Nature}. Moreover they provide as well an experimental realizable laboratory for the study of the as yet undetected Hawking radiation since they are supposed to emit its acoustic counterpart and it is possible to measure their temperature,   and surface gravity with current available techniques \cite{aco1,aco2,aco3}. Their analogue electromagnetic properties have been established as well in \cite{mr1}. Once again theoretical techniques for locating their event horizons are playing an important role in such analyses. 

The purpose of this note is  to provide an algorithm answering  the question whether the horizon of lower dimensional black holes can  be located as the zero of an appropriate curvature invariant. The location of a black hole event horizon can be searched in a non-local way by integrating the equations for the  geodesic motion \cite{wald}. Other possible techniques require approximating the horizon as a minimal Riemannian surface or as a marginally outer trapped surface \cite{asrev}. Recently it has been proved  that it can be located by solving the zeroes of some invariants of the Riemann tensor \cite{sho1, conformal}, and later the same argument has been extended to the use of the Cartan curvature invariants  whose utility has been explicitly demonstrated for (3+1)- and (4+1)-dimensional black holes  \cite{cartan} leading to the {\it geometric horizon conjecture} \cite{geoconj, geoconj2}, which has been tested as well in multi-black holes configurations such as the Kastor-Traschen metric \cite{kastor}. This literature has also shown that those curvature invariants can be used for extracting locally the values of the mass, electric charge and angular velocity of a black hole,  whose mutual relationships can then be compared with the ones predicted by the previously mentioned mechanical laws governing a black hole, and in particular the scaling relation between mass and temperature,   which for our lower dimensional case is going to zero when the former is vanishing contrary to the case of ordinary (3+1)-dimensional black holes \cite{carlip}.
Thus,  it is natural to hypothesize that  this is the procedure we are looking for. The interpretation of the modifications we must apply to this procedure can be used for providing new  insights into the nature of the horizon of a black hole. In fact, the basic ingredient of this method, the curvature, can be decomposed into two parts, only one of which can be determined   by   imposing the Einstein equations. The other is completely free in (3+1)-dimensions, but on the other hand is completely constrained to vanish in lower dimensional frameworks. Thus, a comparison between our techniques for locating the horizon can indeed provide further information on the nature of black holes in different dimensions.   

Our manuscript is organized as follows: in the next section we will review the basic properties of a black hole horizon, in the third section we will motivate more explicitly our interest in a lower dimensional theory, in section four we will  comment on the peculiar properties of (2+1)-dimensional gravity focusing on the consequences of applying our method, while in the fifth section  we will present some explicit applications to the linearly and non-linearly charged Banados-Teitelboim-Zanelli (BTZ) black hole  which generally depends on five physical parameters. These examples of applications clarify the applicability of the method even for a dynamical configuration currently employed in the description of the formation of a black hole.  Then we will move to the same analysis but in (1+1)-dimensional gravity in section six. Finally we will  discuss a connection between the Cartan invariants we constructed for locating the horizon and the behavior of tidal forces on its proximity in section seven, and we conclude in section eight summarizing the differences between our analysis and the case of   (3+1)- and (4+1)-dimensional spacetimes.

\section{Basic properties of the black hole horizon: a very short review}

In this section we will review the basic definitions dealing with black horizons for the sake of completeness.

{\it Definition.} Given a spacetime manifold $\mathcal{M}$, any submanifold which is a spatial section of $\mathcal{M}$ which is topologically a 2-sphere and a null surface, whose any normal has vanishing expansion, and the projection of the stress-energy tensor along any of its null future-directed normal is future-causal, is said to be a \emph{non-expanding horizon} \cite{asrev}.

We notice that such requirements remain true after rescaling the null normals considered. This is the same as for the Killing horizons, which are null hypersurfaces defined by the vanishing of the norm of the time-translational Killing vector field,  in which it is possible to rescale the Killing vector field without affecting it.

However the previous definition does not provide any practical technique for identifying these submanifolds once the full spacetime manifold is provided, that is they do not come from the solution of any differential equation. With this problem in mind, as a first attempt it has been proved that black hole horizons are minimal surfaces, i.e. surfaces which locally minimize their area \cite{minimal1,minimal2}. In this way it was possible to determine that their mean curvature vanishes identically and that they can be obtained from the solution of the Lagrange partial derivative equation derived from an action principle \cite{lagrange}. It is worth noticing that now these are purely geometric methods, while instead the matter content was present in the original definition of the black hole horizon through the stress-energy tensor entering the Einstein equations. Subsequently stationary and static black hole horizons have been shown to be marginally outer trapped surfaces,  which are defined as closed surfaces  on which outward-pointing light rays moving inwards are converging, their expansion being negative. They remain a good approximation even for dynamical configurations \cite{seno}. It is worth noticing that instead this criterion is based on the solution of the Raychauduri equation which depends on the matter content generating the spacetime under examination through its stress-energy tensor.

More recently the localization of the black hole horizon has been connected to the vanishing of certain curvature invariants of the Riemann tensor \cite{sho1,cartan}. Curvature invariants are scalar quantities defined as the traces of products of the curvature tensors, and in general of their derivatives. They are a useful technique for classifying known solutions of the Einstein field equations in reference to the so-called equivalence problem, in which one has to check whether two very different looking solutions (derived in different coordinate systems) are in fact one and the same \cite{coley1}. One of their limitations   is that they cannot distinguish the vanishing scalar invariant (VSI) spacetime from Minkowski \cite{page}.  Specifically, in this paper we are interested in locating the stationary horizon of a black hole which is a null hypersurface orthogonal to a Killing vector field that is null on it. The rigorous   mathematical formulation of the methods we will invesigate  has been proposed in  \cite{sho1,cartan}.
However these theorems constitute only a formal result   on the existence of a curvature invariant which detects the stationary horizon  without delivering the explict curvature invariant which is required, nor providing a constructive algorithm. 
 It is therefore necessary to check  by hands  how they behave for the lower-dimensional black holes, which is the subject of our manuscript. It is interesting to notice again that in this algorithm geometrical and physical properties of the black hole are mixed within the Riemann tensor which can be decomposed into the Weyl tensor  accounting for the former and the Ricci tensor and Ricci scalar accounting for the latter.

\section{Why lower dimensional gravity and black holes?}

Relationships between (2+1)-dimensional gravity with other theoretical frameworks have been explored since its first appearance, the Chern-Simons theory being the most famous example. In this case the dynamics is  derived from an action principle in which the Lagrangian is provided by the Chern-Simons 3-form, meaning that it belongs to the class of Schwarz-type quantum field theory because its observables  are computed exploiting a metric-independent action functional \cite{ati}.  Moreover it admits connections to many other branches of physics ranging from condensed matter theory to quantum gravity. In the former case it was shown that the monopoles exploited for solving the dynamics are related to quantum vortices which typically are topological defects exhibited by superfluids and superconductors.  Its duality with respect to M-theory on $\text{AdS}_4\times \text{S}^7$ has been explored as well \cite{cond}. In the latter case a supergravity theory has been formulated in which supersymmetry and general relativity are combined in a unique model by imposing supersymmetry to fulfill the locality requirement; its availability as effective low-energy limit of string theory has been discussed \cite{super1,super2,super3,super4,super5}. An interesting aspect for our current analysis is that string theory admits a formulation in term of the so-called Born-Infeld Lagrangian, which can be interpreted as representing a non-linear version of the Maxwell equations of electromagnetism \cite{string1,string2,string3}. In fact it has been proved that it exhibits a twofold application in gravity: on one hand it may account for the late time accelerated expansion of the Universe through the Chaplygin gas \cite{bento}, while on the other, solutions picturing charged black holes can be found exactly in this theory as well \cite{born1,born2},   providing an interesting arena for checking whether the current ways of localizing a black hole horizon remain valid.
Applying this way of reasoning with the goal of treating the mathematical non-linearities arising in general relativity,  dilaton-type gravity models have been developed by  Jackiw and Teitelboim \cite{dila}, and by   Callan-Giddings-Harvey-Strominger (CGHS in short)  \cite{CGHS1} exploring different possibilities of coupling between the dilaton field and the one time and one space dimensions of gravity. These theories present  black hole type solutions despite the fact of allowing only flat geometries.

\section{On the use of curvature invariants for detecting a black hole horizon: the peculiar case of lower dimensional Einstein theory}

In this section we want to establish whether the argument proposed in \cite{cartan} for locating the horizon applies also to lower dimensional solutions, and/or if it requires any modification due to the different number of independent components of the curvature tensor. In fact, it is well known that the Weyl tensor identically vanishes for any spacetime in (2+1)-dimensional gravity, implying that the Riemann tensor $R_{\mu\nu\rho\sigma}$ can be fully written 
in terms of the Ricci tensor $R_{\mu\nu}$, the Ricci scalar $R$ and the metric tensor $g_{\mu\nu}$ as follows \cite{carlip}:
\begin{equation}
\label{decom}
R_{\mu\nu\rho\sigma}\, = \, g_{\mu\rho}R_{\nu\sigma} +  g_{\nu\sigma}R_{\mu\rho} - g_{\nu\rho}R_{\mu\sigma} - g_{\mu\sigma}R_{\nu\rho}- \frac12 (g_{\mu\rho}g_{\nu\sigma} - g_{\mu\sigma} g_{\nu\rho}   )R \,.
\end{equation}
The ingredients of (\ref{decom}) must be obtained from the Einstein equations
\beq
\label{efe}
R_{\mu\nu} - \frac12 g_{\mu\nu} R= \Lambda g_{\mu\nu} + T_{\mu\nu}\,,
\eeq
where $\Lambda$ is the cosmological constant, $ T_{\mu\nu}$ is the stress-energy tensor and units such that $8\pi G = c =1$ are used.

Thus the Einstein field equations imply that in the absence of matter-energy content, $ T_{\mu\nu}=0$, the components of the Riemann tensor are constant. Explicitly, we have
\begin{equation}
R_{\mu\nu\rho\sigma}\, = \, \left[2 ( g_{\mu\rho}g_{\nu\sigma} +  g_{\nu\sigma}g_{\mu\rho} - g_{\nu\rho}g_{\mu\sigma} - g_{\mu\sigma}g_{\nu\rho}) - 3 (g_{\mu\rho}g_{\nu\sigma} - g_{\mu\sigma} g_{\nu\rho}   )\right] \Lambda \,.
\end{equation}
Using the Leibniz rule for the covariant derivative, and the fact that the covariant derivative of the metric tensor and of the cosmological constant vanish, we get 
\beq
R_{\mu\nu\rho\sigma ; \eta }\, \equiv \, 0 \,,
\eeq
where a semicolon denotes a covariant derivative, implying that both the scalar polynomial (Karlhede's) invariant ${\mathcal I}= R_{\mu\nu\rho\sigma ; \eta}R^{\mu\nu\rho\sigma ; \eta}$, where repeated indices are summed, and the Cartan invariant ${\mathcal J}= R_{\mu\nu\rho\sigma ; \eta}$ identically vanish in any (2+1)-dimensional spacetime point, and therefore they \emph{cannot} be used for determining the horizon location. 

Note in particular that this claim does not depend on the frame used for computing ${\mathcal J}$, and of course it still holds in the limit $\Lambda \to 0$. Iterating this procedure and increasing the order of differentiation we cannot produce any non-zero higher-order curvature invariant. Moreover using the relationship $g_{\mu\nu} g^{\mu\rho}= \delta_{\nu}{}^{\rho}$, where $\delta_{\nu}{}^{\rho}$ denotes the Kronecker delta, we can also see that the scalar polynomial invariants constructed by the self-contraction of the Riemann tensor, like $R_{\mu\nu\rho\sigma}R^{\mu\nu\rho\sigma}$,
$ R^{\mu\nu\alpha\beta}  R_{\alpha\beta\rho\sigma}R_{\mu\nu}{}^{\rho\sigma}$, etc..., are constant and thus they do not provide any information concerning the horizon location. 
The same conclusion holds trivially even when the invariants ${\mathcal I}$ and ${\mathcal J}$ are computed in terms of the Weyl tensor rather than of the Riemann tensor. This implies that the argument proposed in \cite{cartan}, contrary to the cases of $(3+1)$- and higher dimensional gravity theories, cannot be used for locating the horizon of the corresponding lower dimensional Schwarzschild, Kerr, and Schwarzschild-(anti-)de Sitter black holes because the geometry does not admit enough degrees of freedom. The question now becomes if instead it remains a useful algorithm when the black hole under examination carries an electrical charge which provides a non-trivial Ricci tensor. 

Exploiting the same arguments as in the previous computations, we get
\begin{eqnarray}
\label{cov1}
R_{\mu\nu\rho\sigma ; \eta}&=&  g_{\mu\rho} \left(T_{\nu\sigma ; \eta} -\frac12 g_{\nu\sigma} T_{, \eta}  \right)   +  g_{\nu\sigma} \left(T_{\mu\rho ; \eta} -\frac12 g_{\mu\rho} T_{, \eta}  \right) - g_{\nu\rho}\left(T_{\mu\sigma ; \eta} -\frac12 g_{\mu\sigma} T_{, \eta}  \right) \nonumber\\
&& - g_{\mu\sigma}\left(T_{\nu\rho ; \eta}  - \frac12 g_{\nu\rho} T_{, \eta}  \right) -\frac12 (g_{\mu\rho}g_{\nu\sigma} - g_{\mu\sigma} g_{\nu\rho}   ) \left( -\frac12 T_{, \eta}  \right) \nonumber\\
&=&   g_{\mu\rho} T_{\nu\sigma ; \eta}   +  g_{\nu\sigma} T_{\mu\rho ; \eta}  - g_{\nu\rho}  T_{\mu\sigma ; \eta} 
  - g_{\mu\sigma}T_{\nu\rho ; \eta}  -\frac14 (g_{\mu\rho}g_{\nu\sigma} - g_{\mu\sigma} g_{\nu\rho}   )  T_{, \eta}                     \,,
\end{eqnarray}
where $T$ denotes the trace of the stress-energy tensor entering the Einstein equations. With some algebraic manipulations we get
\beq
{\mathcal I}=4 (T_{\nu\sigma; \eta } T^{\nu\sigma; \eta }+ T_{, \eta } T^{, \eta })\,.
\eeq
 This computation may play a role in providing further restrictions on the matter content the configuration  must obey for getting a useful non-identically zero scalar curvature invariant ${\mathcal I}$.

\section{Explicit cases in 2+1 dimensions}

\subsection{A first explicit example: the case of the  BTZ black hole with electric charge}

For making the reasoning of the  previous section more transparent  we apply it to the test solution (2.14) of \cite{as}:
\begin{eqnarray}
\label{metric}
\d s^2 &=& -N^2(r) \d v^2 +\frac{2r}{K(r)}\d v\d r + K^2(r) (\d\phi +N^\phi (r) \d v)^2, ~~\text{where} \\
N^2(r) &=&  \frac{r^2}{K^2(r)} \left( \frac{r^2}{l^2} - \frac{{\bar l}^2}{2 \pi l^2} Q^2 \ln \frac{r}{r_0} \right),  \\
N^\phi (r) &=& -\frac{\omega}{2\pi K^2(r)} Q^2 \ln \frac{r}{r_0},  \\
K^2(r) &=& r^2 + \frac{\omega^2}{2\pi } Q^2 \ln \frac{r}{r_0}, \\
{\bar l}^2  &=&  l^2 - \omega^2, 
\end{eqnarray}
which describes a (2+1)-dimensional black hole written in Eddington-Finkelstein coordinates ($v$, $r$, $\phi$) with angular velocity $\omega$, electric charge $Q$ and negative cosmological constant $\Lambda=-1/l^2$, while $r_0$ is a reference length related to the mass $M$ of the black hole for making the argument of the logarithm  dimensionless. As pointed out in \cite{as},  the black hole horizon is given by the surface $N(r)=0$, which explicitly provides the solution
\beq
\label{hor}
r_{\rm h}= \frac{Q \sqrt{(\omega^2 - l^2) W\left( \frac{4 \pi r_0^2}{Q^2 (\omega^2 - l^2)} \right)}}{2 \sqrt\pi }\, , 
\eeq
where $W(x)$ denotes the Lambert function which is the inverse function of $x e^x$.

An explicit computation gives us the result
\beq
\label{inv2}
{\mathcal I}= \frac{5 Q^4(l+\omega)^2 (l-\omega)^2   }{2(\pi r^2 l^2)^3} \, \left(2\pi r^2 + Q^2 (\omega^2  -l^2) \ln \frac {r}{r_0}  \right) \,.
\eeq
Solving the equation ${\mathcal I}=0$ we get one and only one solution $r=r_{\rm h}$ implying that the scalar polynomial invariant (\ref{inv2}) detects the black hole horizon without false positives. 

For computing the Cartan invariant ${\mathcal J}$ we fix a ``null'' triad coframe following \cite{segre}
\begin{eqnarray}
l^a&=& \frac {N(r)}{\sqrt2}  \d v,  \\
n^a&=&  \frac {N(r)}{\sqrt2}  \d v - \frac {\sqrt2 r}{N(r) K(r)}  \d r,   \nonumber\\
m^a&=&  \frac {K(r) N^\phi(r)}{\sqrt2}  \d v + \frac {K(r)}{\sqrt2}  \d\phi ,  \nonumber
\end{eqnarray}
such that 
\beq
\label{tr3}
m_a m^a =1 =- l_a n^a \,, \qquad l_a l^a = n_a n^a=l_a m^a=n_a m^a=0 \,,
\eeq
in terms of which the metric (\ref{metric}) reads
\beq
\label{metricframe}
g_{ab}= -2 l_{(a} n_{b)} +2 m_{(a} m_{b)} \,,
\eeq
where round parentheses denote symmetrization. 
 For eliminating possible ambiguities in the construction of the coframe we employ for computing the following Cartan invariants,   we begin by recalling that  the metric tensor is invariant under Lorentz transformations. Therefore (\ref{metricframe})
is invariant under boosts
\beq
l^a \to C_1 l, \qquad n^a \to \frac{1}{C_1}n^a, \qquad m^a \to m^a
\eeq
and null rotations
\beq
l^a \to \frac{1}{2}l^a, \qquad n^a \to \frac{1}{2}C_2^2 l^a+\frac{1}{2}n^a+C_2m^a, \qquad m^a \to C_2l^a+m^a\,,
\eeq
where $C_1$ and $C_2$ are arbitrary functions of the manifold coordinates. Thus we must fix the triad in such a way that the curvature is in its canonical form, i.e. with the smallest possible number of non-zero independent components, which can be done applying the Newman-Penrose formalism. 3-dimensional spacetimes admits  at most 5 independent Ricci scalars defined as \cite{segre}
\begin{eqnarray}
\Phi_{00}&=&\frac{1}{2}R_{ab} l^a l^b\,, \\
\Phi_{22}&=&\frac{1}{2}R_{ab} n^a n^b \,, \\
\Phi_{10}&=&\frac{1}{2\sqrt{2}}R_{ab} m^a l^b \,, \\
\Phi_{12}&=&\frac{1}{2 \sqrt{2}}R_{ab} m^a n^b \,, \\
\Phi_{11}&=&\frac{1}{6}  (R_{ab} m^a m^b +R_{ab} n^a l^b)\,.
\end{eqnarray}
Then, we construct (\ref{tr3}) setting the appropriate components to zero or equal (or proportional to each other) following eqs (5.44)-(5.51) of \cite{segre} according to the isotropy group of the black hole in hand. This procedures is the lower dimensional equivalent to appendix B of \cite{stewart}.

The only functionally independent frame component of the covariant derivative of the Riemann  tensor reads
\beq
\label{inv3}
{\mathcal J}= R_{2113;2}= \frac {\sqrt2 \omega Q^2}{4 l^2r^2\pi} \cdot \frac {Q^2(\omega^2 - l^2)\ln \frac{r}{r_0} + 2r^2\pi}{Q^2\omega^2 \ln \frac{r}{r_0}+2r^2\pi}\,.
\eeq
Setting ${\mathcal J}=0$ we get the only solution $r=r_{\rm h}$,   implying that the above mentioned Cartan invariant constitutes   an appropriate quantity for identifying the black hole horizon. 

We stress the fact once again that ${\mathcal J}$ is computationally less expensive to compute than ${\mathcal I}$ because it is a first degree quantity. Moreover we observe that eqs (\ref{inv2}) and (\ref{inv3}) provide
\beq
\label{limit}
\lim_{Q \to 0}{\mathcal I} \, =\, 0 \,=\, \lim_{Q \to 0}{\mathcal J}
\eeq
confirming the general discussion presented in the previous section.

Finally, we stress that under a $SO(1,1)$ isotropy group,   the frame curvature computed using the Newman-Penrose formalism admits only one independent component  \cite{segre}:
\begin{eqnarray}
\Phi_{11} &=&  \frac16 (R_{ab} m^a m^b + R_{ab} n^a l^b) \,.
\end{eqnarray}
 Note that we are indeed under such a situation since the metric admits two vector fields related to its stationarity and axis-symmetry (as easily recognized from its independence from the coordinates $v$ and $\phi$).

\subsection{Second explicit example: non-linearly charged BTZ black hole}

The metric for a non-linearly charged BTZ black hole is obtained by implementing the Born-Infeld theory within the Einstein equations. It reads \cite{born1}:
\begin{eqnarray}
\d s^2 &=& -f(r) \d t^2 + \frac{\d r^2} {f(r)} + r^2 \d\theta^2 \\
f(r)&=&  \frac{r^2} {l^2} -m +2 r^2 \beta^2 (1-\Gamma) +q^2 \left( 1-2 \ln \frac{r(1+\Gamma)}{2l} \right) \\
\Gamma(r) &=& \sqrt{1+ \frac{q^2}{r^2 \beta^2}}\,,
\end{eqnarray}
where the mass $m$ enters  the metric   directly, and the limit of a charged black hole in the Maxwell theory is obtained in the limit $\beta \to +\infty$, while the effects of the non-linearities become stronger for small $\beta$. Using  polar coordinates we can complement the previous section in which the Eddington-Finkelstein coordinates were adopted instead. The location of the horizon is given by $f(r_{\rm hor})=0$, which can be written implicitly as:
\beq
m=\frac{r_{\rm hor}^2} {l^2} +2 r_{\rm hor}^2 \beta^2 (1-\Gamma(r_{\rm hor})) +q^2 \left( 1-2 \ln \frac{r_{\rm hor}(1+\Gamma(r_{\rm hor}))}{2l} \right)\,.
\eeq
By an explicit computation we obtain
\begin{eqnarray}
{\mathcal I}&=&  \frac{f(r)}{r^4} \cdot [ r^4 (f'''(r))^2  + 4 (r^2 +1) (f''(r))^2    - 8r f''(r) f'(r)  ]  \\
{\mathcal J}&=&  R_{r\theta tr; \theta} \,=\, \frac{\sqrt{f(r)}}{\sqrt2 r^2} [r f''(r) -f'(r)]    \,,
\end{eqnarray}
where the latter was computed with respect to the coframe
\beq
n^a=\frac{\sqrt{f(r)}}{\sqrt2}\d t+\frac{\d r}{\sqrt{2f(r)}}\,, \qquad l^a=\frac{\sqrt{f(r)}}{\sqrt2}\d t-\frac{\d r}{\sqrt{2f(r)}}\,, \qquad m^a=r \d\theta \,.
\eeq
By direct inspection we see that ${\mathcal I}=0={\mathcal J}$ for $f(r)=0$, i.e. they properly detect the horizon. It is straightforward to note as well that they vanish only at the horizon by looking at the quantity within the square brackets which reads 
\begin{eqnarray}
&&\frac{384 \beta^2 q^4 h_1(r) h_2(r)} {(\beta^2 r^2+q^2)^{7/2} (r\beta+\sqrt{\beta^2 r^2+q^2})^6}, ~~~\text{where} \\
&&h_1(r)= \frac54 r^4\beta^4+2 r^2 q^2\beta^2+q^4, ~~\text{and} \nonumber\\
&&h_2(r)=\left(\frac{16}{3} r^6\beta^6+8 q^2 r^4\beta^4+3 q^4 r^2 \beta^2+\frac{q^6}{6}\right)\sqrt{\beta^2 r^2+q^2}+(4\beta^2 r^2+q^2)\beta(\beta^2 r^2+q^2)\left(\frac43 r^2\beta^2+q^2\right)r \nonumber
\end{eqnarray}
for ${\mathcal I}$ and
\begin{eqnarray}
\frac{4(2r^2\beta^2+2\beta r \sqrt{\beta^2 r^2+q^2}+q^2)q^2\beta}{(r\beta+\sqrt{\beta^2 r^2+q^2})^2 \sqrt{\beta^2 r^2+q^2}}
\end{eqnarray}
 for ${\mathcal J}$ which are positive definite. It is clear again from these expressions that (\ref{limit}) still holds. In the strongly non-linear limit we get respectively:
\begin{eqnarray}
{\mathcal I}&\simeq& \frac{64 q^2 \beta^2 }{r^4 l^2}\left[ 2 q^2 l^2 \ln \frac{2\beta l}{q}+(q^2-m)l^2+r^2 \right] +O(\beta^3) \quad {\rm for} \beta\to 0 \\
{\mathcal J}&\simeq& \frac{2 \sqrt{2} |q| \sqrt{q^2 l^2 \ln(4\beta^2 l^2/q^2)-m l^2+r^2
}  }{l r^2}\beta - \frac{4\sqrt{2} q^2 l }{r \sqrt{q^2 l^2 \ln(4\beta^2 l^2/q^2)-m l^2+r^2
} }\beta^2 +O(\beta^3) \quad {\rm for} \beta\to 0\,. \nonumber
\end{eqnarray}

\subsection{The most  general solution in terms of five parameters}

After the warming up of the previous sections, we can   apply our method to   \cite{fivep} which delivers  an original metric in terms of five physical parameters $M$, $J$, $\Lambda$,  $q_\alpha$, and $q_\beta$, which must be regarded as more general than the ones applied in the previous sections.  In a coordinate system ($t$, $r$, $\phi$) it reads:
\begin{eqnarray}
\d s^2 &=& [-N^2(r) +r^2 N_1^2(r)  ] \d t^2 + \frac{\d r^2}{K(r)}+r^2   \d\phi^2 -2 r^2  N_1(r) \d t \d\phi ~~~\text{where}\\
N(r)&=& q_\alpha \left( \frac{J^2}{2 r^2} -M \right) r+ q_\beta \sqrt{  K(r) }, \\
N_1(r) &=&  J\left( \frac{q_\beta}{2 r^2} + \frac{q_\alpha}{r}  \sqrt{  K(r) } \right),\\
K(r)&=& \frac{J^2}{4r^2}-M-\Lambda r^2\,.
\end{eqnarray}
An explicit computation delivers:
\beq
{\mathcal I} = K(r) \cdot {\mathcal F}_1\,,
\eeq
where ${\mathcal F}_1$ is a non-zero function of the metric coefficients, their first and second derivatives. However without the need of expanding any further the computation, we can see that this scalar polynomial invariant vanishes when $K(r)= g^{rr}$ vanishes.
Introducing the coframe
\beq
l^a= \frac{N(r)+r N_1(r)}  {\sqrt{2}} \d t -\frac{r} {\sqrt{2}} \d\phi \,, \qquad n^a= \frac{N(r) - r N_1(r)}  {\sqrt{2}} \d t +\frac{r} {\sqrt{2}} \d\phi \,, \qquad m^a =\frac{1}{\sqrt{2 K(r)}} \d r
\eeq
in terms of which the metric is written as (\ref{metricframe}) we can compute:
\beq
\label{tidal1}
{\mathcal J} = R_{trtt; \phi}=  K^{\frac32}(r) \cdot {\mathcal F}_2 \,,
\eeq
where ${\mathcal F}_2$ is another non-zero function of the metric coefficients, their first and second derivativatives. The same consideration as for the scalar polynomial invariant applies.

\subsection{Curvature invariants and the dynamical formation of a black hole}
\label{secdin}

The cases we studied so far can only account for static and stationary configurations, and thus they assume the existence of the black hole without entering into the details on how it was formed. For addressing this latter problem we will   apply our algorithm to   \cite{dyn12}  which   provides   the solution for a dynamically evolving black hole in (2+1)-dimensional gravity. In this section we will see how the framework provided by the curvature invariants   allows one to follow the mechanism of formation of a black hole (i.e. when in time and where in space a singularity and a horizon appear). Using an Eddington-Finkelstein-like coordinates system ($u$, $r$, $\psi$) the metric reads as follows:
\begin{eqnarray}
\d s^2 &=& -f(u,r) \d u^2 +2 \d u \d r +g(u,r) \d \psi^2, ~~\text{where} \\
f(u,r)&=& \frac{r^2}{l^2}+\frac{8\alpha\left( \tanh \left(\frac{12 \alpha u}{q}  \right)^2-1\right )r}  {q  \tanh \left(\frac{12 \alpha u}{q}  \right)} -\frac{12 \alpha}{q^2}-\frac{\alpha}{q^3 r} \tanh \left(\frac{12 \alpha u}{q}  \right), ~~\text{and}\\
\label{eqg}
g(u,r) &=&  r^2 \tanh \left(\frac{12 \alpha u}{q}  \right)^{\frac23}\,.
\end{eqnarray}
Introducing the coframe
\beq
l^a = \sqrt{ \frac{f(u,r)}{2} }\d u \,, \qquad   n^a = \sqrt{ \frac{f(u,r)}{2}}\d u- \sqrt{ \frac{2}{f(u,r)} } \d r \,, \qquad    m^a = \sqrt{ \frac{g(u,r)}{2} }\d\psi  \,, 
\eeq
in terms of which the metric is written as (\ref{metricframe}),   we can compute:
\beq
{\mathcal J} = R_{r\psi rr; \psi} = \frac{ \sqrt{2} f^{\frac32} (u,r)}{g^3(u,r)}\cdot \left[ g^2(u,r) \frac{\partial^3 g(u,r)} {\partial r^3} - 2 g(u,r) \frac{\partial^2 g(u,r)} {\partial r^2}  \frac{\partial g(u,r)} {\partial r} +\left( \frac{\partial g(u,r)} {\partial r}  \right)^3\right]\,.
\eeq
We  note that this Cartan invariant vanishes for $g^{rr}=f(u,r)=0$ detecting the horizon. Interestingly, substituting eq. (\ref{eqg}) into the square brackets we get zero as well. This seems to be a result following the dynamical property of this solution.

On the other hand, for this metric the scalar polynomial invariant ${\mathcal I}$ does not seem a good tool for locating the horizon.  This is most easily recognized considering a numerical counterexample. For the case $l=1$, $u=5$, $q=2$, $\alpha=-3$, and $r\simeq 0.041659$ we get $f(u,r)=0$, but ${\mathcal I}=-4880+1024i$, where $i$ is the imaginary unit.\footnote{ We stress that for spacetimes that are not stationary, one should search for scalar polynomial curvature invariant that not only vanishes on the horizon of the stationary metric,  but that also remains zero on the horizon under a conformal transformation of the metric \cite{mcconformal}. }

\section{Black holes in 1+1 dimensions}

Grumiller et al.  \cite{1plus1} reviews some interesting black hole solutions in (1+1)-dimensional gravity theory known under the name of Jackiw$-$Teitelboim theory, which are exact solutions in the dilatonic gravity theory. If we adopt the coordinate system ($u$, $r$), the general line element we must consider is:
\beq
\d s^2= -\xi(r) \d u^2 +2 \d r \d u \,.
\eeq
Relevant  subcases are:
\begin{itemize}
\item A  Schwarzschild-like solution   written in the chiral gauge  \cite{sc1plus1}:
\beq
\xi(r)= 1-\frac{2M}{r}
\eeq
\item A  Reissner-Nordstr\"om-like black hole   which constitutes a generalization of the previous solution which accounts for an electric charge:
\beq
\xi(r)= 1-\frac{2M}{r}+\frac{Q^2}{r^2}
\eeq
\item  A black hole in an asymptotically  Minkowski spacetime in the Eddington-Finkelstein gauge:
\beq
\label{generala}
\xi(r)= 2 C_0 |1-a|^{\frac{a}{a-1}} r^{\frac{a}{a-1}} \left( \frac{B}{a} \right)^{\frac{2-a}{2(a-1)}} +1
\eeq
\item   A black hole in an asymptotically  Rindler spacetime in the Eddington-Finkelstein gauge:
\beq
\xi(r)= r B_2^{\frac12} - M  r^{\frac{a}{a-1}}
\eeq
\item   A black hole in an asymptotically  (anti)-de Sitter spacetime in the Eddington-Finkelstein gauge:
\beq
\xi(r)= r^2 B - M  r^{\frac{a}{a-1}}
\eeq
\item A black hole in CGHS gravity  whose asymptotic region is $r\to -\infty$\footnote{The CGHS black hole corresponds to the case with $a=1$, and therefore a different system of coordinates is required for dealing with the singularity that occurs in (\ref{generala}).}:
\beq
\xi(r)= 1+\frac{2C_0 e^{\sqrt{B} r}} {B} \,.
\eeq
\end{itemize}
The  latter solutions picture black holes in asymptotically Minkowski, Rindler and (anti)-de Sitter spaces respectively. In particular we can recognize the linear and  quadratic terms in the fourth and fifth cases which are characteristic for Rindler and de Sitter backgrounds.  We note that in this class of theories the Weyl tensor is again trivially identical to zero, thus we will work in terms of the Riemann tensor. A straightforward computation delivers:
\beq
{\mathcal I} = \left( \frac{ \text{d}^3 \xi(r)}{\d r^3}   \right)^2 \xi(r)\,.
\eeq
Thus this scalar polynomial invariant correctly identifies the Killing horizon, i.e. it vanishes where $\xi(r)=0$. 
Intoducing the co-frame
\beq
l^a = \sqrt{ \frac{\xi(r)}{2}  } \d u\,, \qquad  n^a = \sqrt{ \frac{\xi(r)}{2}  } \d u - \sqrt{ \frac{2}{\xi(r)}  } \d r  \,,
\eeq
in terms of which the metric reads as
\beq
g_{ab}= -2 l_{(a}  n_{b)} \,,
\eeq
we can compute the Cartan invariant
\beq
\label{tidal2}
{\mathcal J} =R_{ruru;u} = \frac{\sqrt{2 \xi(r)}}{4} \cdot \frac{\text{d}^3 \xi(r)}{\d r^3} , 
\eeq
which detects the killing horizon for the same reason stated above.
Furthermore, we note that the fact of having or not having false positives depends on the  behavior of the third derivative of the non-trivial metric coefficient, which will assume different values depending on the particular solution considered. For the cases reported in this paper we get, respectively,
\begin{eqnarray}
\frac{\text{d}^3 \xi(r)}{\d r^3}  &=& \frac{12M}{r^4}, \\
\frac{\text{d}^3 \xi(r)}{\d r^3}  &=& \frac{12M}{r^4} -  \frac{24 Q^2}{r^5}, \\
\frac{\text{d}^3 \xi(r)}{\d r^3}  &=&  -  \frac{  2 C_0 (a-2) |a-1|^{\frac{a}{a-1 }} r^{\frac{3-2a}{a-1}}  B^{\frac {2-a}{2(a-1)}}    a^{\frac {3a -4 }{2(a-1)}}  }    {(a-1)^3 },     \\
\frac{\text{d}^3 \xi(r)}{\d r^3}  &=& \frac{Ma(a-2)  r^{\frac{3-2a}{a-1}} }{(a-1)^3 },  \\
\frac{\text{d}^3 \xi(r)}{\d r^3}  &=& \frac{Ma(a-2)  r^{\frac{3-2a}{a-1}} }{(a-1)^3 },   \\
\frac{\text{d}^3 \xi(r)}{\d r^3}  &=& 2C \sqrt{B} e^{r \sqrt{B}}.
\end{eqnarray}
Thus a false positive is possible in the third, fourth and fifth case for the particular choice of the solution parameter $a=2$. The fourth and fifth cases vanish as well for $r=0$ which is another root of $\xi(r)=0$ for the corresponding black holes. Moreover, a false positive can be seen also in the second case,   in the particular case $M=\frac{2\sqrt{3}Q}3$ after substituting $r_{\rm hor}=M \pm \sqrt{M^2 -Q^2}$.

Grumiller et al.  \cite{1plus1} discusses as well  the following  dynamical solution:
\beq
\d s^2= 2 \d r \d u +K(r,u) \d u^2 \,.
\eeq
A direct computation shows that
\beq
\label{pol1+1} 
{\mathcal I} = \frac{\partial^3 K(r,u)}{\partial r^3} \left( 2\frac{\partial^3 K(r,u)}{\partial u \partial r^2} -K(r,u)\frac{\partial^3 K(r,u)}{\partial  r^3} \right)\,.
\eeq
Defining
\beq
l^a=\left(\frac{1}{\sqrt2} -1 \right) \sqrt{K(r,u)} \d u\,, \qquad  n^a=\left(\frac{1}{\sqrt2} +1 \right) \sqrt{K(r,u)} \d u + \frac{\sqrt2}{(\sqrt2 -1)\sqrt{K(r,u)}} \d r  \,,
\eeq
it is possible to compute
\beq
\label{car1+1} 
{\mathcal J} =R_{rrur;r} = \frac{\sqrt2 (\sqrt2 -1)   \sqrt{K(r,u)} }{4} \cdot  \frac{\partial^3 K(r,u)}{\partial r^3}\,,
\eeq
which vanishes for $K(r,u)=0$. Recalling our previous discussion about static (1+1)-dimensional black holes, whose generalization is considered here, we can exclude the second factor of this Cartan invariant to vanish in general, meaning that its only zeroes are indeed the zeroes of the metric coefficient $K(r,u)$.
Thus we can appreciate a difference between the two methods. When we use the scalar polynomial invariant (\ref{pol1+1}) we need further assumptions on the explicit form of the metric coefficient $K(r,u)$, information that are not required when we try to locate the horizon as the zero of a Cartan curvature invariant (\ref{car1+1}).

Of some interest to us as well is the black hole solution arising from the Callan-Giddings-Harvey-Strominger model \cite{1plus1, CGHS1}:
\beq
\d s^2 = G(u,v) \d u \d v\,, \text{where} \qquad G(u,v)=- \left( \frac{M}{\lambda} -\lambda^2 uv\right)^{-1}, 
\eeq
written  in the coordinate system ($u$, $v$). A direct computation shows that
\beq
{\mathcal I} =8 \lambda^3 uv M^2 G^3(u,v),
\eeq
which vanishes for $G(u,v)=0$. Fixing the coframe
\beq
l^a=-\frac{\sqrt2}2G(u,v) \d u\,, \qquad n^a=\frac{\sqrt2}2 \d v
\eeq
we can compute:
\begin{eqnarray}
{\mathcal J}&=&R_{vuvv;u}= 2 \sqrt{2}  \frac{ G(u,v) \left[  G(u,v) \frac{\partial^3 G(u,v)}{\partial v^2 \partial u} - \frac{\partial^2 G(u,v)}{\partial v^2 }   \frac{\partial G(u,v)}{\partial u} -3 \frac{\partial^2 G(u,v)}{\partial v \partial u} \frac{\partial G(u,v)}{\partial v}     \right] +3 \frac{\partial G(u,v)}{\partial u} \left( \frac{\partial G(u,v)}{\partial v} \right)^2  }{G^4(u,v)}   \nonumber \\
&=&  -2 \sqrt{2} \lambda^3 uM G^2(u,v)   \,,
\end{eqnarray}
where in the last equality we have used the explicit form of the metric coefficient. Thus this seems to vanish only after an appropriate choice of the metric. A false positive is present for $u=0$.

\section{Physical interpretation of the horizon-detecting Cartan invariants}

 The tidal tensor is defined as:
\beq
E^{i}{}_{j}=\frac{\partial \Phi}{\partial x^i \partial x^j}
\eeq
where $\Phi$ is the gravitational potential. Tidal forces affect the geodesic deviation equation for the displacement vector $\xi^i$   via
\beq
\frac{{\rm d}^2 \xi^i}{{\rm d} t^2}=-E^{i}{}_{j}\xi^j\,,
\eeq
allowing us  to rewrite the tidal tensor in terms of the curvature Riemann tensor as \cite{schutz}
\beq
E^{i}{}_{j}=R^{i}{}_{tjt}
\eeq
where $t$ denotes the time index. Therefore we can interpret two of the Cartan invariants we used for locating the horizon (\ref{inv3})-(\ref{tidal1}) as providing a local extremum with respect to a spatial direction of the tidal  force. The same interpretation can be proposed for (\ref{tidal2})  but with respect to  time,   suggesting  that the black hole (i.e. its horizon) forms when a local extremum of the tidal forces is reached. Taking into account that Cartan invariants are foliation-independent \cite{cartanfr} this seems to suggest that the horizon   can be physically characterized by a local extremum of the tidal forces. Therefore a hypothetical spaceship crossing the horizon should {\it in principle} be able to recognize it from a change in the behavior of the measured tidal force.  We propose  that this behavior of the tidal forces can be a valuable technique for distinguishing black holes from other  star-like objects which do not have an horizon and therefore according to \cite{cartan} do not exhibit this property of the Cartan invariants (that is having a spacetime point in which a Cartan invariant connected to the tidal tensor vanishes).

 Interests   in lower-dimensional black holes is motivated from the existence of physical systems whose motion is known to be confined in lower dimensions, like cosmic strings and domain walls which arise in the Polyakov model \cite{LL1,LL2} (see also \cite{LL3} for a review about applications of BTZ black holes in the light of this connection). The existence of hydrodynamic equilibrium has led to the proposal of stellar   models in (1+2) gravity \cite{LL4,LL5,LL6}. Distinguishing a lower-dimenional black hole from a lower-dimensional star is important in laboratory settings which investigate Bose-Einstein condensate \cite{LL7},  and the optical properties of graphene \cite{LL8} through analogue sonic BTZ black holes. The variations of the tidal forces detected using  absolute gravimeters \cite{LL9},  and interpreted following this section can provide an answer to this issue.

\section{Discussion and Conclusion}

When we consider a (3+1)-dimensional Schwarzschild-like metric in spherical coordinates ($t$, $r$, $\theta$, $\phi$)
\beq
\d s^2= -f(r) \d t^2 + \frac{\d r^2}{f(r)}+r^2 \d\theta^2 + r^2 \sin^2 \theta \d\phi^2 \,,
\eeq
we can compute straightforwardly the two curvature invariants
\begin{eqnarray}
{\mathcal I}&=& \frac{f(r)}{r^6}  \left[ r^6 (f'''(r))^2 + 8 r^4 (f''(r))^2 - 16 r^3 f''(r) f'(r) + 16 r^2 (f'(r))^2 -32 r (f(r)-1) f'(r)+32 (f(r)-1)^2     \right]      \nonumber\\
{\mathcal J}&=& \frac{\sqrt{f(r)}}{2 \sqrt2 r^2} \left[ f'(r)  -r f''(r)       \right]      \,,
\end{eqnarray}
where a prime denotes a derivative with respect to $r$, and the latter invariant has been computed with respect to the frame 
\beq
l^a = \frac{\sqrt{f(r)}}{\sqrt2} \d t - \frac{1}{ \sqrt{2f(r)}} \d r, \quad n^a = \frac{\sqrt{f(r)}}{\sqrt2} \d t + \frac{1}{ \sqrt{2f(r)}} \d r, \quad m^a = \frac{r}{\sqrt2} \d\theta + \frac{i r \sin\theta}{\sqrt2 } \d\phi, \quad {\bar m}^a = \frac{r}{\sqrt2} \d\theta - \frac{i r \sin\theta}{\sqrt2 } \d\phi,
\eeq
where $i$ denotes the imaginary unit.
We note that in such a case ${\mathcal I}=0={\mathcal J}$ where $f(r)=0=g^{rr}$. This implies that these two curvature invariants locate the horizon of the black hole without invoking any assumption concerning the Einstein field equations, that is they work for the Schwarzshild solution, the Schwarzshild  (anti-)de Sitter metric, the Schwarzschild-NUT and Reissner-N\"ordstrom black holes, etc. Interestingly they provide information about the horizon location also for vacuum spacetimes. 
This is a consequence of the non-triviality of the Weyl tensor in 3+1 dimensions. 

On the other hand, when we move to lower dimensional (2+1)-dimensional theory the Weyl tensor vanishes identically and the Riemann tensor becomes a constant in vacuum and for spacetimes characterized only by the cosmological constant. An even more drastic phenomenon takes place in the gravitational theories characterized by only one temporal and one spatial  dimensions since their geometries are all necessarily flat. Thus we cannot take for guaranteed that this argument holds  automatically for detecting the location of lower-dimensional black hole horizons. The purpose of this work was to investigate qualitatively and quantitatively how the  dimensionality of the theory for gravity adopted influence the values of the curvature invariants which have been used for locating the horizon of a black hole in \cite{cartan}.  In particular,  we examined in detail the role of the matter content checking explicitly  that when we are dealing with a non-vacuum solution the applicability of the  general method is recovered because of the non-triviality of the Riemann curvature. Moreover in section  (\ref{secdin}) we discussed the reasons why the Cartan curvature invariant should be regarded as a more powerful method than the polynomial one for tracing the horizon of a dynamical black hole.

 \cite{wald} explicitly states the necessity of knowing the entire spacetime evolution for claiming if a black hole is present or not. Therefore this is why our technique is important: it does not require all this amount of information. We do not need to evolve any initial data, making this computationally convenient when numerical relativity studies are adopted. Belonging to the same line of thinking,   \cite{gravitation} compares and contrasts local and global aspects of black hole horizons,  explaining   that they are located by propagating light rays. Again, our method is intended to provide a reliable alternative.
Also the state of the art of numerical techniques still relies on (expensive) non-local techniques: see for example
 \cite{bentivegna,clifton,yoo,jhep}  where in particular the last one   focuses on horizon deformations in lower dimensional geometries by numerical methods in the context of the AdS/CFT correspondence.  To this purpose \cite{shibata} investigates how new numerical relativity techniques can locate horizons already known in the literature. Our research intends  to complement it from an analytical perspective.   The research looking for connections between the zeros of certain curvature invariants and the location of the horizon has begun because numerical relativity simulations can extract the mass and angular momentum of a black hole only by locating its horizon and by computing its area through the excision method. Finding the location of the black hole horizon provides information about which region must be excised  \cite{lake}.

It is curious to see how our research fits inside the broader context dealing with the determination of a black hole horizon shortly reviewed in the second section. The very first technique for finding the horizon was based on the completely geometrical concept of minimal surface; then the subsequent technique based on the ideas of marginally outer trapped surfaces and curvature invariants involved a mixture of physical and geometrical properties of the black holes, while now we have shown that in lower dimensional gravity the black hole horizon seems to be a purely physical object because only information about the matter content of the spacetime were used for locating it arriving at a conceptually different method for finding it than the one we started from. In  light of this  we have considered both the Maxwell theory of electromagnetism and the non-linear Born-Infeld theory for the modeling of the matter content in (2+1)-dimensional gravity, as well as the case of a dilatonic field coupled to gravity in 1+1 dimension.

Information about the curvature of a given spacetime are encoded into its Riemann tensor. A clearer description can be obtained by decomposing it into its trace, the Ricci tensor, and its trace-free part, the Weyl tensor. While the former is fully determined once information  of  the matter content is provided via the Einstein equations, the latter remains  in part free because  its value is not provided by the field equations, but only some constraints must be accounted for through the Ricci identities. Thus we can intuitively speak of the Ricci tensor as the ``physical" part of the curvature, and of the Weyl tensor as its ``geometrical" part. This difference can be better appreciated by the textbook example of the triplet of Minkowski, Schwarzschild and Kerr metrics, which are all vacuum solutions in general relativity, meaning that they have the same Ricci tensor, but they have different Weyl tensors, which in turn makes their full curvature different, with the Minkowski spacetime being flat and the other two not.  This can be appreciated also considering cosmological solutions: one can have a dust Friedman model and a dust Bianchi I, Bianchi V, or Bianchi IX models, with same matter content (dust, i.e. same physics entering the Einstein equations) but different geometries (Weyl part of the curvature). However   this decomposition of the Riemann tensor does not hold anymore in lower dimensional gravity because the Weyl tensor is completely determined by the number of dimensions of the theory and it is identically zero. Thus, in our language the curvature is a purely ``physical" property of the spacetime because the Riemann tensor can be fully eliminated in terms of the stress-energy tensor as we used in eq. (\ref{cov1}), and even all solutions in (1+1)-dimensions are flat due to stricter restrictions. Once we have established that the horizon of a black hole can be found exploiting some quantities derived from the spacetime curvature, as studied in this paper, the previous decomposition of the Riemann tensor can provide a new characterization the properties of the horizon, and in particular it turns out that in lower dimensional theories it is reduced to a purely ``physical" property, in our language, of the solutions, because its curvature is fully determined by the field equations.

\begin{acknowledgments}
Y.C.O.  thanks NNSFC (grant No.11705162) and the Natural Science Foundation of Jiangsu Province
(No.BK20170479) for funding support.
\end{acknowledgments}

{}

\end{document}